%% file: version1.tex
\documentclass[aps, prl, reprint]{revtex4-1}

\pdfoutput=1 
\usepackage{graphicx}
\usepackage{color,amsmath}
\usepackage[normalem]{ulem}

\usepackage[colorlinks,linkcolor=blue,citecolor=blue,urlcolor=blue]{hyperref}
\usepackage{siunitx}
\usepackage{comment}
\usepackage{layouts}

\begin{document}

\title{Phase Slips and Parity Jumps in Quantum Oscillations of Inverted InAs/GaSb Quantum Wells}

\author{Matija Karalic}
\email{makarali@phys.ethz.ch}
\affiliation{Solid State Physics Laboratory, ETH Zurich, 8093 Zurich, Switzerland}

\author{Christopher Mittag}
\affiliation{Solid State Physics Laboratory, ETH Zurich, 8093 Zurich, Switzerland}

\author{Susanne Mueller}
\affiliation{Solid State Physics Laboratory, ETH Zurich, 8093 Zurich, Switzerland}

\author{Thomas~Tschirky}
\affiliation{Solid State Physics Laboratory, ETH Zurich, 8093 Zurich, Switzerland}

\author{Werner Wegscheider}
\affiliation{Solid State Physics Laboratory, ETH Zurich, 8093 Zurich, Switzerland}

\author{Leonid Glazman}
\affiliation{Department of Physics, Yale University, New Haven, Connecticut 06520, USA}

\author{Klaus Ensslin}
\affiliation{Solid State Physics Laboratory,  ETH Zurich, 8093 Zurich, Switzerland}

\author{Thomas Ihn}
\affiliation{Solid State Physics Laboratory, ETH Zurich, 8093 Zurich, Switzerland}

\date{\today}

\begin{abstract}
We present magnetotransport measurements of a strongly hybridized inverted InAs/GaSb double quantum well. We find that the spin-orbit interaction leads to an appreciable spin-splitting of hole-like states, which form distinct Landau levels in a perpendicular magnetic field. The resulting quantum Hall state is governed by a periodic even and odd total filling arising due to the simultaneous occupation of electron-like and hole-like Landau levels of differing degeneracy. Furthermore, oscillatory charge transfer between all involved subbands leads to discrete phase slips in the usual sequential filling of Landau levels, and coincidentally the phase slips are close to $\pi$. These results shed new insights on the Landau level structure in composite systems and have consequences for interpreting intercepts obtained from index plots, which are routinely employed to investigate the presence of Berry's phase.

\end{abstract}

\maketitle

InAs/GaSb double quantum wells (QWs) are a composite semiconductor system that hosts spatially separated electrons and holes and exhibits significant spin-orbit interaction (SOI). The thicknesses of the constituent InAs and GaSb QWs determine the band alignment, which may be either inverted or noninverted. Application of electric and magnetic fields offers additional tunability, allowing for continuous control of this band alignment \cite{naveh_bandstructure_1995, qu_electric_2015, suzuki_gate-controlled_2015}. In the inverted phase, electrons and holes hybridize, opening an energy gap \cite{lakrimi_minigaps_1997, yang_evidence_1997, cooper_resistance_1998}, which facilitates the formation of the quantum spin Hall insulator (QSHI) state under the right conditions \cite{liu_quantum_2008, knez_evidence_2011, suzuki_edge_2013, knez_observation_2014, mueller_nonlocal_2015, du_robust_2015}.

The inverted band structure and tunability of InAs/GaSb double QWs enable many interesting experiments not directly related to the QSHI state, such as the manipulation of the SOI in single- and two-carrier regimes \cite{beukman_spin-orbit_2017}, the observation of a giant spin-orbit splitting close to the charge neutrality point (CNP) \cite{nichele_giant_2017} and tunable mixing of quantum Hall (QH) edge states \cite{karalic_lateral_2017}.

Here, we report our findings on the Landau level (LL) structure in a strongly inverted InAs/GaSb double QW featuring enhanced hybridization. Using transport measurements we uncover a periodic even and odd filling of Landau levels leading to a checkerboard pattern in the longitudinal resistivity $\rho_{xx}$. Additionally, an anomalous shift violates the usual $1/B_\perp$-periodic sequence of LL filling. By analyzing Shubnikov-de Haas (SdH) oscillations and performing two and three-band transport modeling, we unravel the electron and hole-like charge carrier distribution in the system, thereby deducing the presence of an initially concealed SOI split hole-like subband. We then explain how the combination of SOI induced splitting and Landau quantization can lead to an unconventional filling sequence of LLs in this coupled electron-hole bilayer. 

\begin{figure}[!t]
\includegraphics[width=\columnwidth]{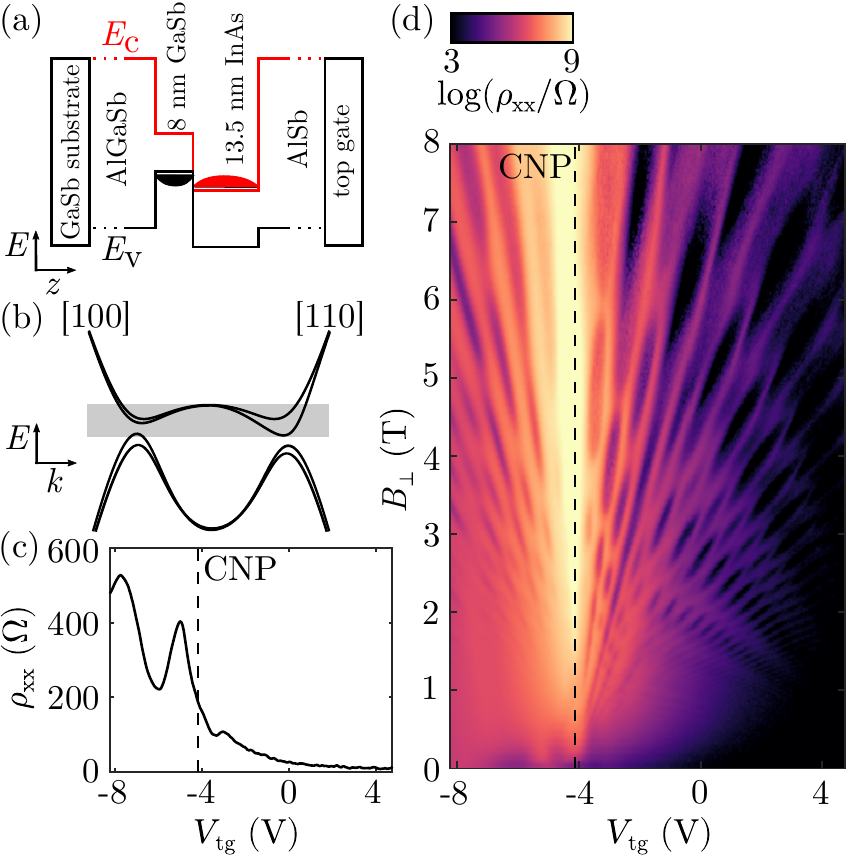}
\caption{\textbf{(a)} Conduction (valence) band edge energy $E_\mathrm{c}$ ($E_\mathrm{v}$) as a function of growth direction $z$ for the InAs/GaSb system. \textbf{(b)} Schematic band structure $E(k)$ of the double QW system. The shaded area is investigated in subsequent experiments. \textbf{(c)} $\rho_{xx}$ at $B_\perp = 0$ as function of $V_\mathrm{tg}$ with the position of the CNP highlighted. \textbf{(d)} $\rho_{xx}$ as function of $V_\mathrm{tg}$ and $B_\perp$ with the position of the CNP highlighted.}
\label{fig1}
\end{figure}

Measurements were performed on a gated Hall bar of $\SI{10}{\micro\meter}$ width and $\SI{20}{\micro\meter}$ length oriented along the $[0\bar{1}1]$ crystallographic direction on a heterostructure consisting of an $\SI{8}{\nano\meter}$ GaSb QW and a $\SI{13.5}{\nano\meter}$ InAs QW [Fig.\,\ref{fig1}(a)]. The Hall bar was defined by wet etching and covered by a SiN layer separating the Ti/Au top gate from the heterostructure surface. Ohmic contacts were fabricated by etching through the gate dielectric, then selectively down to the InAs QW and depositing Ti/Au. All measurements were conducted in a dilution refrigerator at a base temperature of $\SI{135}{\milli\kelvin}$ using low-frequency lock-in techniques with constant ac current bias.

The inverted band structure of our coupled double QWs is schematically depicted in Fig.\,\ref{fig1}(b). The thickness of the InAs QW is sufficiently large to drive the system into the semimetallic phase. This implies enhanced hybridization between electron and hole bands, while the anisotropy of the dispersion effectively quenches the hybridization gap. The Rashba-type SOI already present in the constituent QWs is amplified by the hybridization, and leads to significant spin-splitting of valence and conduction bands \cite{zakharova_strain-induced_2002, li_spin_2009, Hu_electric_2016}. The longitudinal resistivity $\rho_{xx}$, see Fig.\,\ref{fig1}(c), shows no local resistance maximum at or close to the CNP, indicating the lack of a true energy gap, as expected. A resistance maximum gradually evolves at the CNP upon applying a perpendicular magnetic field $B_\perp$. In the shaded area in Fig.\,\ref{fig1}(b) between the CNP and the former top of the GaSb valence band both electrons (majority) and holes (minority charge carriers) are present. This region is probed in the following.

Figure\,\ref{fig1}(d) shows a map of $\rho_{xx}$ as a function of top gate voltage $V_\mathrm{tg}$ and $B_\perp$. The voltage $V_\mathrm{tg}$ tunes the total charge carrier density in the system. The CNP divides the map into two regions with electrons (holes) being the majority charge carriers to the right (left) of the CNP. Two sets of lines fanning outwards from the CNP follow minima in $\rho_{xx}$ in both regions. Remarkably, we discern an atypical yet regular pattern in the distribution of minima. Minima in $\rho_{xx}$ of constant filling are modulated, moving towards and away from zero resistivity in a systematic fashion depending on their position in the ($V_\mathrm{tg}$, $B_\perp$) parameter space. We now focus on the region to the right of the CNP where electrons are in the majority and discuss the properties and origins of the observed checkerboard pattern. 

\begin{figure}[!t]
\includegraphics[width=\columnwidth]{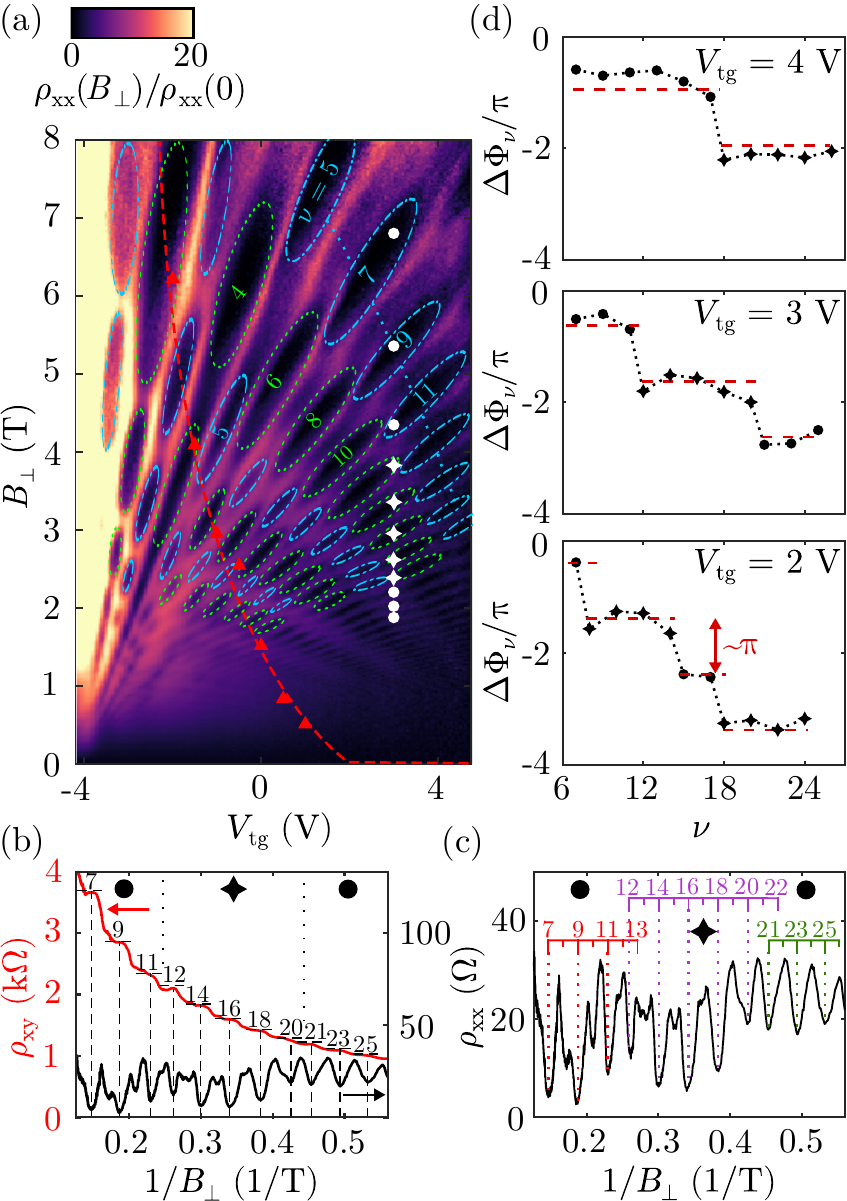}
\caption{\textbf{(a)} Zoom-in of Fig.\,\ref{fig1}(d). Contour lines marking minima associated with either even or odd total filling factor $\nu$ are colored differently, with $\nu$ given explicitly for some minima. The dotted line connecting several minima of odd parity exemplifies how adjacent  minima of the same parity lie on lines of negative slope. Circles (stars) mark positions of odd $\nu$ (even $\nu$) minima at $V_\mathrm{tg} = \SI{3}{\volt}$ as in (b), (c), (d). Triangles and the dashed line serving as a guide to the eye correspond to the situation where the missing density $p_2$, as introduced in the main text, equals $e B_\perp /h$. \textbf{(b)} $\rho_{xx}$, $\rho_{xy}$ as function of $1/B_\perp$ at $V_\mathrm{tg} = \SI{3}{\volt}$ with positions of minima in $\rho_{xx}$ and expected positions of plateaus in $\rho_{xy}$, $h/ie^2$ with integer $i$, marked by dashed lines. The associated $\nu = i$ are also indicated. \textbf{(c)} $\rho_{xx}$ reproduced from (b) with minima marked by dotted lines and rulers associated with each segment of constant parity of $\nu$ showing the expected positions of minima and the affiliated values of $\nu$. \textbf{(d)} Steps in the oscillations' phase as a function of $\nu$ for several values of $V_\mathrm{tg}$, see the main text for details. Horizontal dashed lines spaced by $\Delta\Phi_\nu = \pi$ provide a guide to they eye.} 
\label{fig2}
\end{figure}

The region of interest in Fig.\,\ref{fig1}(d) is reproduced in Fig.\,\ref{fig2}(a) for clarity. Figure.\,\ref{fig2}(b) is a cut at constant total density at $V_\mathrm{tg} = \SI{3}{\volt}$, showing both longitudinal and transverse resistivities $\rho_{xx}$ and $\rho_{xy}$ versus $1/B_\perp$. Well-developed plateaus in $\rho_{xy}$ are described by $\rho_{xy} = h/\nu e^2$ with integer $\nu$ and occur concomitantly with minima in $\rho_{xx}$. Because we probe a composite system with multiple charge carrier species, $\nu$ corresponds to a total filling factor, taking both electron and hole-like LLs into account. While the minima typically do not reach zero, the quantization in $\rho_{xy}$ suffices for an unambiguous assignment of $\nu$. Looking at the sequence of plateaus in $\rho_{xy}$, we deduce that $\nu$ decreases in increments of two with $1/B_\perp$, with the exception of selected transitions indicated by the dotted lines. There, $\nu$ changes by one only. In this way, the parity of $\nu$ switches between even (denoted by stars) and odd (circles) as a function of magnetic field. The positions of the minima in $\rho_{xx}$ at $V_\mathrm{tg} = \SI{3}{\volt}$ are  also marked by symbols in Fig.\,\ref{fig2}(a). The minima in $\rho_{xx}$ corresponding to the missing plateaus are suppressed. Using the quantization of $\rho_{xy}$, we assign a filling factor to each minimum and highlight minima of even and odd $\nu$ in Fig.\,\ref{fig2}(a) with differently colored contour lines. We observe that neighboring minima of the same parity seemingly lie on lines of negative slope, as illustrated by the dotted line. A cut at fixed $V_\mathrm{tg}$ such as in Fig.\,\ref{fig2}(b) typically intersects multiple such lines, so that minima correspond to $\nu$ being piecewise even or odd. The pattern appears to change upon approaching the CNP, becoming more complex. Similar even-odd behavior in few-layer transition metal dichalcogenides (TMDCs) was attributed to interplay between cyclotron and Zeeman energies as well as to a density dependent $g$-factor originating from interaction effects \cite{movva_density-dependent_2017, larentis_large_2018, pisoni_interactions_2018}. We also observed weak signs of said behavior in InAs/GaSb double QWs in the semiconductor (topological insulator) phase, attributing it to avoided crossings between LLs mediated by ordinary and spin-orbit interband coupling effects \cite{karalic_experimental_2016}.

In addition to the unconventional filling sequence discussed above, there exists another peculiarity in the form of discrete phase slips occurring whenever the parity switches. Figure.\,\ref{fig2}(c) depicts $\rho_{xx} (1/B_\perp)$, again at $V_\mathrm{tg} = \SI{3}{\volt}$. Starting from low $1/B_\perp$, we see that the minima corresponding to $\nu =7, 9, 11$ are equidistantly spaced in $1/B_\perp$. However, the subsequent $\nu =12$ minimum is not located at the expected position, but instead halfway between where the $\nu = 12$ and $\nu = 13$ would lie according to the periodicity set by the $\nu =7, 9, 11$ minima. The same phenomenon repeats itself at higher $1/B_\perp$ at the transition from even to odd $\nu$. There, the $\nu = 21$ minimum lies halfway between the nonexistent $\nu = 21$ and  $\nu = 22$ minima which would follow the periodicity determined by the observed $\nu = $ 12--20 minima. The period in $1/B_\perp$ remains approximately constant regardless of the shifts. We have verified that the shifts occur generically for all ($V_\mathrm{tg}$, $B_\perp$) shown in Fig.\,\ref{fig2}(a) whenever the parity of $\nu$ changes.

In general, we may describe the shifts in terms of discrete phase slips. To quantify the phase slips, we extract an average total density $n_\mathrm{QHE}$ of charge carriers in the QH state by piecewise fitting of $\nu (1/B_\perp)$ in an index plot for fixed $V_\mathrm{tg}$ \cite{supp}. Then, we calculate the phase slip $\Delta \Phi_\nu /\pi$ using $\Delta \Phi_\nu /\pi = 2h n_\mathrm{QHE} \Delta (1/B_\perp)/e$, where $\Delta (1/B_\perp)$ is the difference in $1/B_\perp$ between the expected position of the minimum corresponding to $\nu$, $e \nu /h n_\mathrm{QHE}$, and the position where it actually occurs. Figure\,\ref{fig2}(d) presents the evolution of $\Delta \Phi_\nu /\pi$ for several $V_\mathrm{tg}$. The phase is seen to jump downwards by around $\pi$ whenever the parity switches.

\begin{figure}[!t]
\includegraphics[width=\columnwidth]{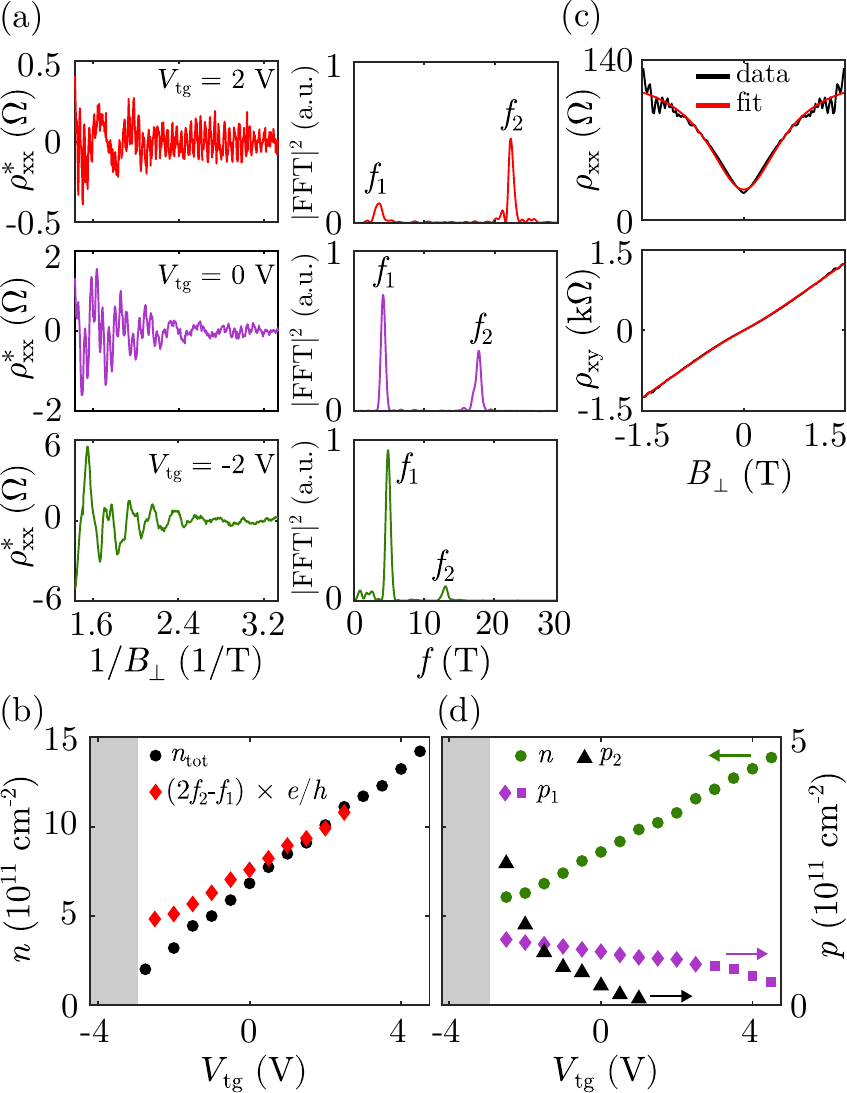}
\caption{\textbf{(a)} Exemplary low-field SdH oscillations after subtraction of the smooth background with the corresponding normalized power spectra for several $V_\mathrm{tg}$. \textbf{(b)} Hall density $n_\mathrm{tot}$ and the total density $(2f_2 - f_1) \times e/h$ obtained from the SdH oscillations as function of $V_\mathrm{tg}$. In the shaded region, no reliable data points are available. \textbf{(c)} Example for the simultaneous fitting of $\rho_{xx}$, $\rho_{xy}$ using a three-band model at $V_\mathrm{tg} = 0$. \textbf{(d)} As in (b), but showing $n$, the electron density, as well as $p_1$ and $p_2$, the densities of the spin-split hole subbands. $p_1$ is determined by both SdH oscillations (diamonds) and two-band fitting (squares).}
\label{fig3}
\end{figure}

To understand the origin of the even-odd periodicity and the phase slips we investigate the densities of all charge carriers by analysis of SdH oscillations and with the help of two and three-band transport models at low fields \cite{supp}, as displayed in Fig.\,\ref{fig3}. The low-field SdH oscillations exhibit a single frequency $f_2$ for $V_\mathrm{tg} > \SI{2.5}{\volt}$ which decreases upon decreasing $V_\mathrm{tg}$ and therefore corresponds to electron-like states [Fig.\,\ref{fig3}(a)]. The frequency $f_2$ is related to the Hall density $n_\mathrm{tot}$ obtained from fitting $\rho_{xy}$ above a certain $B_\perp$-value where $\rho_{xy}$ is linear in $B_\perp$ through $n_\mathrm{tot} \approx 2f_2 \times e/h$. Because $n_\mathrm{tot}$ determined in this way measures the total density of free charge carriers, the factor of two implies that the population imbalance due to spin-splitting is negligible and that carriers essentially populate a twofold degenerate electron-like band. For $V_\mathrm{tg} \leq \SI{2.5}{\volt}$, an additional frequency $f_1$ appears, see Fig.\,\ref{fig3}(a). The frequency $f_1$ increases upon decreasing $V_\mathrm{tg}$, implying hole-like states. Close to $V_\mathrm{tg} = \SI{2.5}{\volt}$ we have $n_\mathrm{tot} \approx (2f_2 - f_1) \times e/h$, signifying that $f_1$ describes a single spin-orbit split hole subband. Upon decreasing $V_\mathrm{tg}$ further a systematic deviation between $n_\mathrm{tot}$ and $(2f_2 - f_1) \times e/h$ appears, as observed in Fig.\,\ref{fig3}(b). The frequencies $f_1$ and $f_2$ cannot account for all charge carriers, increasingly overestimating the density, $n_\mathrm{tot} < (2f_2 - f_1)\times e/h$. This motivates us to look for the second spin-orbit split hole subband containing the missing holes that does not partake in SdH oscillations. We therefore fit $\rho_{xx}$ and $\rho_{xy}$ simultaneously with a three-band model by inverting $\sigma_{xx} = \sum \sigma_{xx}^{i}$ and $\sigma_{xy} = \sum \sigma_{xy}^{i}$ with $\sigma_{xx}^{i}, \sigma_{xy}^{i}$ being the conductivities of the individual bands ($i = 1, 2, 3$). We neglect intersubband scattering and, for consistency, fix two of the three densities to $p_1 = f_1 \times e/h$ and $n = 2f_2 \times e/h$, respectively, leaving four fitting parameters, namely $p_2$, the missing density, and three mobilities $\mu_i$ \cite{supp}. Figure.\,\ref{fig3}(c) depicts the result of such a fit at $V_\mathrm{tg} = 0$. The complete result for all $V_\mathrm{tg}$ is shown in Fig.\,\ref{fig3}(d). The three-band fitting works for $V_\mathrm{tg} \leq \SI{1}{\volt}$. For $V_\mathrm{tg} > \SI{1}{\volt}$ $p_2$ is too small compared to $p_1$ and $n$ to be determined reliably. In fact, for $V_\mathrm{tg} > \SI{1}{\volt}$ a two-band model sufficiently describes the data and for $1 < V_\mathrm{tg} \leq \SI{2.5}{\volt}$ it results in densities that agree with $p_1$ and $n$ as defined above. For $V_\mathrm{tg} > \SI{2.5}{\volt}$ a two-band fit allows us to determine the continuation of $p_1$ where $f_1$ disappears from the SdH oscillations. Note that the two SOI split hole subbands are degenerate at $k = 0$ in the zero field limit due to time reversal symmetry and $p_2 = 0$ is impossible given $p_1 > 0$. While $p_2$ is too small to be detected experimentally for $V_\mathrm{tg} > \SI{1}{\volt}$, it does not vanish completely.

We found that the spin-splitting of the electron-like band of density $n$ cannot be experimentally resolved. The same is true for the conventional Zeeman splitting of this band \cite{note1}. Hole-like states exist in two subbands and have different dispersions due to the SOI, and therefore Landau quantization of each results in nondegenerate levels. The hole-like subband of density $p_1$ enters the QH state, whereas the subband of density $p_2$ does not. Above the dashed line in Fig.\,\ref{fig2}(a), $p_2 < e B_\perp /h$, being insufficient to change the total filling factor by one, and we may think of the corresponding holes as forming a background density. 

\begin{figure}[!t]
\includegraphics[width=\columnwidth]{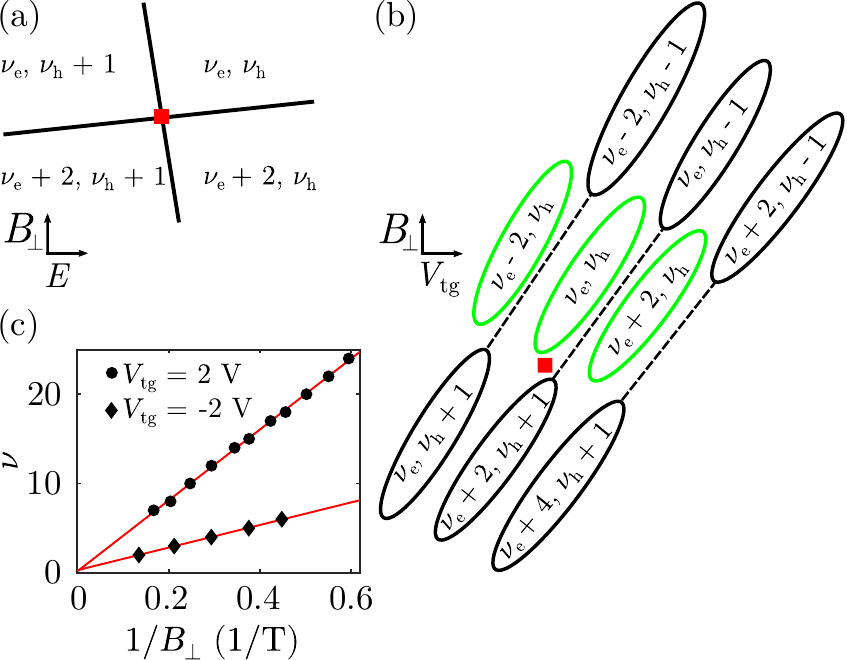}
\caption{\textbf{(a)} Filling of LLs in $(E, B_\perp)$-space in the vicinity of a crossing point between a doubly degenerate electron-like LL and a nondegenerate hole-like LL of corresponding filling $\nu_\mathrm{e}$ and $\nu_\mathrm{h}$, respectively. \textbf{(b)} As in (a), but upon conversion to $(V_\mathrm{tg}, B_\perp)$-space. The crossing point from (a) is now located at the position indicated by the square. Dashed lines connect minima of constant $\nu$. \textbf{(c)} Index plots demonstrating the extrapolation of $\nu$ for $1/B_\perp \to 0$ at $V_\mathrm{tg} = \SI{2}{\volt}$ and $\SI{-2}{\volt}$.} 
\label{fig4}
\end{figure}

The dispersion of LLs in the vicinity of a (anti-) crossing point between twofold degenerate electron-like and nondegenerate hole-like levels is schematically depicted by lines in Fig.\,\ref{fig4}(a) together with $\nu_e$ and $\nu_h$, the filling factors of the electron-like and hole-like levels, respectively ($\nu = \nu_e-\nu_h$). Filling factors $\nu_e$ are even and change in increments of two, whereas $\nu_h$ is even or odd and changes in increments of one. Converting from $(E, B_\perp)$ to $(V_\mathrm{tg}, B_\perp)$, we obtain the diagram in Fig.\,\ref{fig4}(b), recognizing the even-odd pattern from the experiment. The hole mass being larger than the electron mass explains why typically several electron-like LLs are (de)populated before a hole-like LL is (de)populated for constant $V_\mathrm{tg}$.  A simple density of states model illustrating this fact is shown in the Supplemental Material \cite{supp}. 

We now turn to the phase slips and elucidate their source. The Fermi energy oscillates as a function of $B_\perp$ at constant \textit{total} density for fixed $V_\mathrm{tg}$. However, the densities of the individual subsystems, independent of their character, or whether they are in the QH state or not, also oscillate together with the Fermi energy. The phase slips occur because the density of charge carriers in the QH state is not constant due to the self-consistent transfer of charge back and forth between the electron ($n$) and hole-like states ($p_1$) in the QH state and the hole-like states constituting the background ($p_2$). In this picture, the fact that the phase slips are around $\pi$ is coincidence. 

An important consequence of the phase slips is that the intercept $\gamma$ for $1/B_\perp \to 0$ in an index plot $\nu (1/B_\perp)$ is not meaningful. The method for extracting Berry's phase in Dirac materials like graphene or three-dimensional topological insulator surface states from quantum oscillations by $1/B_\perp \to 0$ extrapolation \cite{novoselov_two-dimensional_2005, taskin_berry_2011, wright_quantum_2013, kuntsevich_simple_2018} is \textit{not} applicable here. More generally, in a fully quantized two-dimensional system, $\gamma$ is either $0$ or $1/2$. Performing the extrapolation anyway, see Fig.\,\ref{fig4}(c), we can find a trivial intercept $\gamma \approx 0$ away from the CNP ($\gamma = 0.18 \pm 0.17$ at $V_\mathrm{tg} =\SI{2}{\volt}$) and a nontrivial $\gamma$ close to the CNP ($\gamma = 0.30 \pm 0.05$ at $V_\mathrm{tg} =\SI{-2}{\volt}$). A nontrivial intercept close to the CNP was recently reported in Ref.\,\citenum{nichele_giant_2017}, also in inverted InAs/GaSb double QWs, and understood in terms of Berry's phase. Our results offer an alternative explanation that does not involve Berry's phase and are also relevant for other two-dimensional composite systems with SOI, such as HgTe QWs and TMDCs \cite{konig_quantum_2007, qian_quantum_2014, wu_observation_2018}, which are of fundamental interest for the realization of the QSHI state. 

In conclusion, we have investigated an unconventional LL filling in inverted InAs/GaSb double QWs. Electron-hole hybridization and the SOI lead to an even-odd periodicity upon Landau quantization, resulting in a checkerboard pattern in $\rho_{xx}$ as a function of density and magnetic field that is caused by the difference in degeneracies between electron and hole-like LLs. Additional, clandestine hole-like states that do not directly appear in SdH oscillations and in the QH state nevertheless have a profound impact, leading to abrupt phase slips in the usual $1/B_\perp$-periodic sequence of LLs due to intersubband charge transfer. This phenomenon can lead to nontrivial intercepts obtained from index plots, which are however not associated with a nontrivial Berry's phase. Our findings are not specific to InAs/GaSb double QWs, but are also applicable to other composite systems, for example to systems with appreciable SOI induced splitting. 

\begin{acknowledgments}
The authors acknowledge the support of the ETH FIRST laboratory and the financial support of the Swiss Science Foundation (Schweizerischer Nationalfonds, NCCR QSIT) and thank Andrea Hofmann for input. L.G. acknowledges the support by NSF DMR Grant No. 1603243.
\end{acknowledgments}

%\bibliography{bibl}
\input{bblBibliography.bbl}

\newpage

\section{Supplemental Material}
In this Supplemental Material, we provide additional details pertaining to the analysis presented in the main text. Furthermore, we present a simple calculation of the Landau level dispersion to enable comparison with the experiment.

\setcounter{equation}{0}
\renewcommand{\theequation}{A.\arabic{equation}}
\setcounter{figure}{0}
\renewcommand{\thefigure}{A.\arabic{figure}}

\subsection{Determination of $n_\mathrm{QHE}$}
\begin{figure}[!h]
\includegraphics[width=\columnwidth]{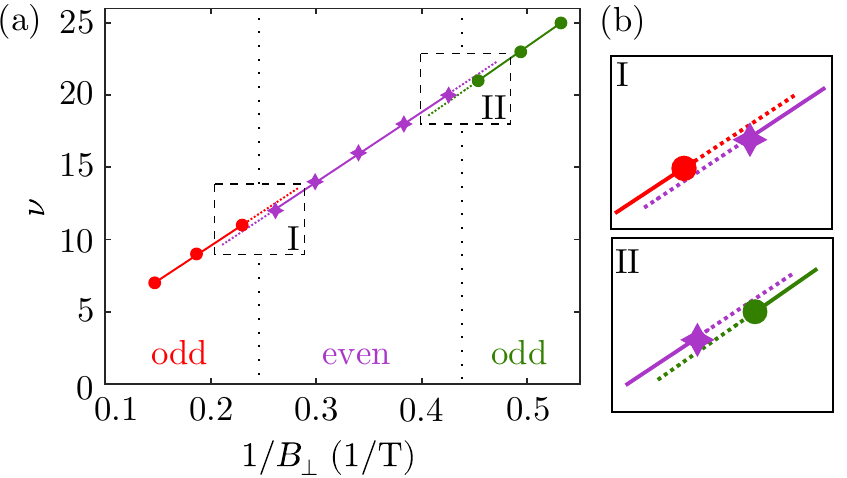}
\caption{\textbf{(a)} Index plot $\nu (1/B_\perp)$ at $V_\mathrm{tg} = \SI{3}{\volt}$. Stars and circles mark points corresponding to even and odd total filling, respectively. Vertical dotted lines demarcate regions of differing parity of $\nu$. Solid lines are linear regression fits to the data points in each region. Dotted lines are extensions of the solid lines that penetrate into adjacent regions. The zoom-ins in \textbf{(b)} show the situation at the boundary of adjacent regions, as indicated in (a) by the dashed rectangles.} 
\label{fig1supp}
\end{figure}

In order to find $n_\mathrm{QHE}$ we turn to index plots $\nu (1/B_\perp)$, such as the one in Fig.\,\ref{fig1supp}(a) which is taken at $V_\mathrm{tg} = \SI{3}{\volt}$, see also Figs.\,\ref{fig2}(b), (c) of the main text. The total filling factor $\nu$ is piecewise even (stars) or odd (circles). In each segment $i$ of constant parity, we fit $\nu (1/B_\perp)$ with a linear function $\nu = a_i/ B_\perp + \gamma_i$. The slope $a_i$ is related to the total density of charge carriers quantized to Landau levels, $n_\mathrm{QHE}^i$, through $n_\mathrm{QHE}^i = a_i \times e/h$. For the case of Fig.\,\ref{fig1supp}(a), we obtain the following result (in order of increasing $\nu$): $a_1 = (48.22 \pm 1.24)$\,T, $a_2 = (48.48 \pm 0.72)$\,T, $a_3 = (51.01 \pm 0.83)$\,T. We then perform an inverse-variance weighting to calculate a weighted average value for the slope, $a$. Finally, we take $n_\mathrm{QHE} =  a \times e/h$. The quantity $n_\mathrm{QHE}$ therefore describes the average total density of charge carriers that are quantized in $B_\perp$. At sufficiently large $B_\perp$ these charge carriers enter the quantum Hall state. For Fig.\,\ref{fig1supp}(a), we find $n_\mathrm{QHE} = (1.19 \pm 0.01)\times 10^{12}$\,cm$^{-2}$.

Fig.\,\ref{fig1supp}(b) depicts zoom-ins of Fig.\,\ref{fig1supp}(a) in the vicinity of where the parity of $\nu$ changes abruptly. The dotted lines indicate extensions of the fits into the adjacent regions of differing parity. We recognize that there is a discontinuity in the form of a horizontal shift between line segments associated with regions of differing parity. This shift is equivalent to the phase slip discussed in the main text. The points in the index plot do not all lie on a single line. 

\subsection{Analysis of Shubnikov-de Haas Oscillations}
The analysis of the Shubnikov-de Haas (SdH) oscillations consists of using the Fourier transform to extract the power spectrum, allowing us to identify the frequencies present in the oscillations.

We start with the longitudinal resistivity $\rho_{xx} (B_\perp)$ and invert the magnetic field axis to obtain $\rho_{xx} (1/B_\perp)$. There is no need for interpolation because the measurement points are acquired equidistantly in $1/B_\perp$ to begin with. Then, we subtract a low-order polynomial (typically of 6th order) from $\rho_{xx} (1/B_\perp)$ to remove the slowly varying background, obtaining $\rho_{xx}^* (1/B_\perp)$. This is followed by multiplying with a Hamming window and zero-padding, before using the fast Fourier transform to determine the discrete Fourier transform. We convert the frequency axis $f$ obtained in this manner to a density axis by multiplication with $e/h$ and take the modulus squared of the complex-valued amplitude to obtain the power spectrum. Finally, we normalize the power spectrum by dividing with the integrated power.

The magnetic field range used in the analysis of the SdH oscillations here is $\SI{0.3}{\tesla} \leq B_\perp \leq \SI{0.7}{\tesla}$.

\subsection{Two- and Three-band Transport Modeling}
In the presence of multiple occupied subbands the classical (Drude) behavior of both longitudinal and transverse resistivities $\rho_{xx}$ and $\rho_{xy}$ is modified with respect to the case of a single subband. In the simplest scenario which assumes no intersubband interactions, the subbands contribute to transport as parallel channels and the conductivities $\sigma_{xx}^{i}$ and $\sigma_{xy}^{i}$ of the individual subbands are summed up to obtain the total conductivities $\sigma_{xx} = \sum \sigma_{xx}^{i}$ and $\sigma_{xy} = \sum \sigma_{xy}^{i}$. $i$ is the subband index. $\sigma_{xx}^{i}$ and $\sigma_{xy}^{i}$ are given by
\begin{equation}
\sigma_{xx}^{i} = \frac{e n_i \mu_i}{1+\mu_i^2 B_\perp^2} 
\label{eq1}
\end{equation}
and
\begin{equation}
\sigma_{xy}^{i} = \pm \frac{e n_i \mu_i^2 B_\perp}{1+\mu_i^2 B_\perp^2}, 
\label{eq2}
\end{equation}
where $n_i$ is the density and $\mu_i$ is the Drude mobility of the $i$-th subband. The sign of $\sigma_{xy}^{i}$ is different for electron-like ($+$) and hole-like ($-$) subbands. $\rho_{xx}$ and $\rho_{xy}$ are calculated according to
\begin{equation}
\rho_{xx} = \frac{\sigma_{xx}}{\sigma_{xx}^2+\sigma_{xy}^2} 
\label{eq3}
\end{equation}
and
\begin{equation}
\rho_{xy} = \frac{\sigma_{xy}}{\sigma_{xx}^2+\sigma_{xy}^2}.
\label{eq4}
\end{equation}

To fit our data with the two- and three-band transport models, we first symmetrize and antisymmetrize  $\rho_{xx}$ and $\rho_{xy}$, respectively. We then fit $\rho_{xx}$ and $\rho_{xy}$ simultaneously with the expressions given above to obtain $n_i$ and $\mu_i$. If some $n_i$ are already known, e.g., from the analysis of SdH oscillations, they can be used as fixed input parameters. We fit iteratively in the interval $0 \leq \lvert B_\perp \rvert \leq \text{1.4--} \SI{1.5}{\tesla}$ and rescale $\rho_{xx}$ with respect to $\rho_{xy}$ according to the square root of the ratio of their residual sum of squares in order to achieve the same or similar errors in both quantities.

\subsection{Calculation of the Landau Level Dispersion}
We carry out basic calculations of the Landau level (LL) dispersion based on a series of simplifications. We assume the presence of a parabolic, doubly degenerate electron-like band described by a constant effective mass $m_\mathrm{e}^*$ and a parabolic, nondegenerate hole-like band described by a constant effective mass $m_\mathrm{h, 1}^*$. These bands represent the bands of densities $n$ and $p_1$ from the main text. They form LLs that have a density of states (DOS) characterized by Gaussian broadening with energies $\Gamma_\mathrm{e}$ and $\Gamma_ \mathrm{h}$, respectively, which scale with $\sqrt{B_\perp}$. The band alignment is inverted in the sense that the bottom of the electron-like band lies below the top of the hole-like band. The band offset is a linear function of the top gate voltage $V_\mathrm{tg}$, decreasing with decreasing $V_\mathrm{tg}$ due to the changing perpendicular electric field (the conductive substrate forming the back gate is always grounded). Zeeman splitting is not taken into account. Apart from the two aforementioned bands, there is an additional band with a steplike DOS that is either zero (above the top of the original hole-like band) or $m_\mathrm{h, 2}^*/(2 \pi \hbar^2)$ (below the top of said band). It corresponds to the hole-like band of density $p_2$.

\begin{figure}[!h]
\includegraphics[width=\columnwidth]{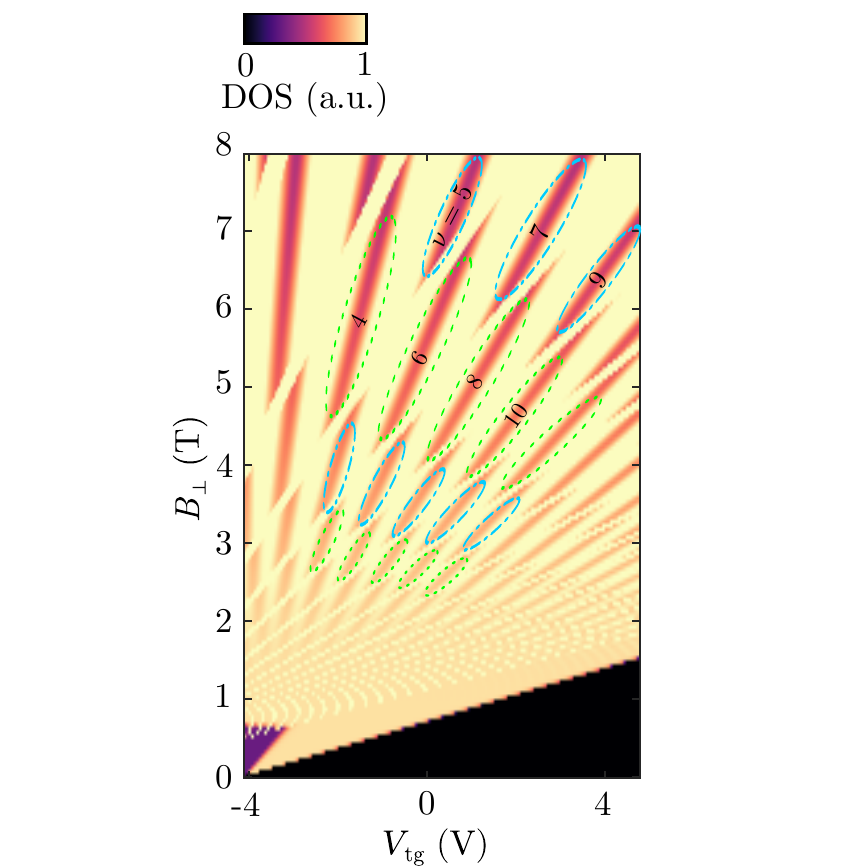}
\caption{Calculated DOS in arbitrary units as function of $V_\mathrm{tg}$ and $B_\perp$. In the dark triangular area on the bottom right, the number of LLs used is insufficient to host all charge carriers and therefore the calculation is invalid in this region. Differently colored contour lines mark minima associated with either even or odd total filling factor $\nu$, with $\nu$ given explicitly for some minima. The parameters used are: $m_\mathrm{e}^* = 0.036 \times m_0$, $m_\mathrm{h, 1}^* = 0.36 \times m_0$, $m_\mathrm{h, 2}^* = 0.18 \times m_0$, $\Gamma_\mathrm{e} = \SI{0.1}{\milli\electronvolt}$, $\Gamma_\mathrm{h} = \SI{0.2}{\milli\electronvolt}$. A total of 20 LLs are taken into account per band.} 
\label{fig2supp}
\end{figure}

To perform the calculation we determine the position of the Fermi energy for given magnetic field $B_\perp$ and total density $n_\mathrm{tot} (V_\mathrm{tg})$. Then, we calculate the total DOS and plot it as a function of $B_\perp$ and $V_\mathrm{tg}$. The result of such a calculation is shown in Fig.\,\ref{fig2supp}. There is satisfactory qualitative agreement with Fig.\,\ref{fig2}(a) of the main text. Minima in the DOS are expected to correspond to minima in $\rho_{xx}$. In Fig.\,\ref{fig2supp}, several total filling factors $\nu$ are indicated for ease of comparison and minima in the DOS are marked by differently colored contour lines in analogy to the main text.

\clearpage

\end{document}

%% file: bblBibliography.bbl

%% file: version1.bbl
\begin{thebibliography}{29}%
\makeatletter
\providecommand \@ifxundefined [1]{%
 \@ifx{#1\undefined}
}%
\providecommand \@ifnum [1]{%
 \ifnum #1\expandafter \@firstoftwo
 \else \expandafter \@secondoftwo
 \fi
}%
\providecommand \@ifx [1]{%
 \ifx #1\expandafter \@firstoftwo
 \else \expandafter \@secondoftwo
 \fi
}%
\providecommand \natexlab [1]{#1}%
\providecommand \enquote  [1]{``#1''}%
\providecommand \bibnamefont  [1]{#1}%
\providecommand \bibfnamefont [1]{#1}%
\providecommand \citenamefont [1]{#1}%
\providecommand \href@noop [0]{\@secondoftwo}%
\providecommand \href [0]{\begingroup \@sanitize@url \@href}%
\providecommand \@href[1]{\@@startlink{#1}\@@href}%
\providecommand \@@href[1]{\endgroup#1\@@endlink}%
\providecommand \@sanitize@url [0]{\catcode `\\12\catcode `\$12\catcode
  `\&12\catcode `\#12\catcode `\^12\catcode `\_12\catcode `\%12\relax}%
\providecommand \@@startlink[1]{}%
\providecommand \@@endlink[0]{}%
\providecommand \url  [0]{\begingroup\@sanitize@url \@url }%
\providecommand \@url [1]{\endgroup\@href {#1}{\urlprefix }}%
\providecommand \urlprefix  [0]{URL }%
\providecommand \Eprint [0]{\href }%
\providecommand \doibase [0]{http://dx.doi.org/}%
\providecommand \selectlanguage [0]{\@gobble}%
\providecommand \bibinfo  [0]{\@secondoftwo}%
\providecommand \bibfield  [0]{\@secondoftwo}%
\providecommand \translation [1]{[#1]}%
\providecommand \BibitemOpen [0]{}%
\providecommand \bibitemStop [0]{}%
\providecommand \bibitemNoStop [0]{.\EOS\space}%
\providecommand \EOS [0]{\spacefactor3000\relax}%
\providecommand \BibitemShut  [1]{\csname bibitem#1\endcsname}%
\let\auto@bib@innerbib\@empty
%</preamble>
\bibitem [{\citenamefont {Naveh}\ and\ \citenamefont
  {Laikhtman}(1995)}]{naveh_bandstructure_1995}%
  \BibitemOpen
  \bibfield  {author} {\bibinfo {author} {\bibfnamefont {Y.}~\bibnamefont
  {Naveh}}\ and\ \bibinfo {author} {\bibfnamefont {B.}~\bibnamefont
  {Laikhtman}},\ }\href {\doibase 10.1063/1.113297} {\bibfield  {journal}
  {\bibinfo  {journal} {Applied Physics Letters}\ }\textbf {\bibinfo {volume}
  {66}},\ \bibinfo {pages} {1980} (\bibinfo {year} {1995})}\BibitemShut
  {NoStop}%
\bibitem [{\citenamefont {Qu}\ \emph {et~al.}(2015)\citenamefont {Qu},
  \citenamefont {Beukman}, \citenamefont {Nadj-Perge}, \citenamefont {Wimmer},
  \citenamefont {Nguyen}, \citenamefont {Yi}, \citenamefont {Thorp},
  \citenamefont {Sokolich}, \citenamefont {Kiselev}, \citenamefont {Manfra},
  \citenamefont {Marcus},\ and\ \citenamefont
  {Kouwenhoven}}]{qu_electric_2015}%
  \BibitemOpen
  \bibfield  {author} {\bibinfo {author} {\bibfnamefont {F.}~\bibnamefont
  {Qu}}, \bibinfo {author} {\bibfnamefont {A.~J.}\ \bibnamefont {Beukman}},
  \bibinfo {author} {\bibfnamefont {S.}~\bibnamefont {Nadj-Perge}}, \bibinfo
  {author} {\bibfnamefont {M.}~\bibnamefont {Wimmer}}, \bibinfo {author}
  {\bibfnamefont {B.-M.}\ \bibnamefont {Nguyen}}, \bibinfo {author}
  {\bibfnamefont {W.}~\bibnamefont {Yi}}, \bibinfo {author} {\bibfnamefont
  {J.}~\bibnamefont {Thorp}}, \bibinfo {author} {\bibfnamefont
  {M.}~\bibnamefont {Sokolich}}, \bibinfo {author} {\bibfnamefont {A.~A.}\
  \bibnamefont {Kiselev}}, \bibinfo {author} {\bibfnamefont {M.~J.}\
  \bibnamefont {Manfra}}, \bibinfo {author} {\bibfnamefont {C.~M.}\
  \bibnamefont {Marcus}}, \ and\ \bibinfo {author} {\bibfnamefont {L.~P.}\
  \bibnamefont {Kouwenhoven}},\ }\href {\doibase
  10.1103/PhysRevLett.115.036803} {\bibfield  {journal} {\bibinfo  {journal}
  {Physical Review Letters}\ }\textbf {\bibinfo {volume} {115}},\ \bibinfo
  {pages} {036803} (\bibinfo {year} {2015})}\BibitemShut {NoStop}%
\bibitem [{\citenamefont {Suzuki}\ \emph {et~al.}(2015)\citenamefont {Suzuki},
  \citenamefont {Harada}, \citenamefont {Onomitsu},\ and\ \citenamefont
  {Muraki}}]{suzuki_gate-controlled_2015}%
  \BibitemOpen
  \bibfield  {author} {\bibinfo {author} {\bibfnamefont {K.}~\bibnamefont
  {Suzuki}}, \bibinfo {author} {\bibfnamefont {Y.}~\bibnamefont {Harada}},
  \bibinfo {author} {\bibfnamefont {K.}~\bibnamefont {Onomitsu}}, \ and\
  \bibinfo {author} {\bibfnamefont {K.}~\bibnamefont {Muraki}},\ }\href
  {\doibase 10.1103/PhysRevB.91.245309} {\bibfield  {journal} {\bibinfo
  {journal} {Physical Review B}\ }\textbf {\bibinfo {volume} {91}},\ \bibinfo
  {pages} {245309} (\bibinfo {year} {2015})}\BibitemShut {NoStop}%
\bibitem [{\citenamefont {Lakrimi}\ \emph {et~al.}(1997)\citenamefont
  {Lakrimi}, \citenamefont {Khym}, \citenamefont {Nicholas}, \citenamefont
  {Symons}, \citenamefont {Peeters}, \citenamefont {Mason},\ and\ \citenamefont
  {Walker}}]{lakrimi_minigaps_1997}%
  \BibitemOpen
  \bibfield  {author} {\bibinfo {author} {\bibfnamefont {M.}~\bibnamefont
  {Lakrimi}}, \bibinfo {author} {\bibfnamefont {S.}~\bibnamefont {Khym}},
  \bibinfo {author} {\bibfnamefont {R.~J.}\ \bibnamefont {Nicholas}}, \bibinfo
  {author} {\bibfnamefont {D.~M.}\ \bibnamefont {Symons}}, \bibinfo {author}
  {\bibfnamefont {F.~M.}\ \bibnamefont {Peeters}}, \bibinfo {author}
  {\bibfnamefont {N.~J.}\ \bibnamefont {Mason}}, \ and\ \bibinfo {author}
  {\bibfnamefont {P.~J.}\ \bibnamefont {Walker}},\ }\href {\doibase
  10.1103/PhysRevLett.79.3034} {\bibfield  {journal} {\bibinfo  {journal}
  {Physical Review Letters}\ }\textbf {\bibinfo {volume} {79}},\ \bibinfo
  {pages} {3034} (\bibinfo {year} {1997})}\BibitemShut {NoStop}%
\bibitem [{\citenamefont {Yang}\ \emph {et~al.}(1997)\citenamefont {Yang},
  \citenamefont {Yang}, \citenamefont {Bennett},\ and\ \citenamefont
  {Shanabrook}}]{yang_evidence_1997}%
  \BibitemOpen
  \bibfield  {author} {\bibinfo {author} {\bibfnamefont {M.~J.}\ \bibnamefont
  {Yang}}, \bibinfo {author} {\bibfnamefont {C.~H.}\ \bibnamefont {Yang}},
  \bibinfo {author} {\bibfnamefont {B.~R.}\ \bibnamefont {Bennett}}, \ and\
  \bibinfo {author} {\bibfnamefont {B.~V.}\ \bibnamefont {Shanabrook}},\ }\href
  {\doibase 10.1103/PhysRevLett.78.4613} {\bibfield  {journal} {\bibinfo
  {journal} {Physical Review Letters}\ }\textbf {\bibinfo {volume} {78}},\
  \bibinfo {pages} {4613} (\bibinfo {year} {1997})}\BibitemShut {NoStop}%
\bibitem [{\citenamefont {Cooper}\ \emph {et~al.}(1998)\citenamefont {Cooper},
  \citenamefont {Patel}, \citenamefont {Drouot}, \citenamefont {Linfield},
  \citenamefont {Ritchie},\ and\ \citenamefont
  {Pepper}}]{cooper_resistance_1998}%
  \BibitemOpen
  \bibfield  {author} {\bibinfo {author} {\bibfnamefont {L.~J.}\ \bibnamefont
  {Cooper}}, \bibinfo {author} {\bibfnamefont {N.~K.}\ \bibnamefont {Patel}},
  \bibinfo {author} {\bibfnamefont {V.}~\bibnamefont {Drouot}}, \bibinfo
  {author} {\bibfnamefont {E.~H.}\ \bibnamefont {Linfield}}, \bibinfo {author}
  {\bibfnamefont {D.~A.}\ \bibnamefont {Ritchie}}, \ and\ \bibinfo {author}
  {\bibfnamefont {M.}~\bibnamefont {Pepper}},\ }\href {\doibase
  10.1103/PhysRevB.57.11915} {\bibfield  {journal} {\bibinfo  {journal}
  {Physical Review B}\ }\textbf {\bibinfo {volume} {57}},\ \bibinfo {pages}
  {11915} (\bibinfo {year} {1998})}\BibitemShut {NoStop}%
\bibitem [{\citenamefont {Liu}\ \emph {et~al.}(2008)\citenamefont {Liu},
  \citenamefont {Hughes}, \citenamefont {Qi}, \citenamefont {Wang},\ and\
  \citenamefont {Zhang}}]{liu_quantum_2008}%
  \BibitemOpen
  \bibfield  {author} {\bibinfo {author} {\bibfnamefont {C.}~\bibnamefont
  {Liu}}, \bibinfo {author} {\bibfnamefont {T.~L.}\ \bibnamefont {Hughes}},
  \bibinfo {author} {\bibfnamefont {X.-L.}\ \bibnamefont {Qi}}, \bibinfo
  {author} {\bibfnamefont {K.}~\bibnamefont {Wang}}, \ and\ \bibinfo {author}
  {\bibfnamefont {S.-C.}\ \bibnamefont {Zhang}},\ }\href {\doibase
  10.1103/PhysRevLett.100.236601} {\bibfield  {journal} {\bibinfo  {journal}
  {Physical Review Letters}\ }\textbf {\bibinfo {volume} {100}},\ \bibinfo
  {pages} {236601} (\bibinfo {year} {2008})}\BibitemShut {NoStop}%
\bibitem [{\citenamefont {Knez}\ \emph {et~al.}(2011)\citenamefont {Knez},
  \citenamefont {Du},\ and\ \citenamefont {Sullivan}}]{knez_evidence_2011}%
  \BibitemOpen
  \bibfield  {author} {\bibinfo {author} {\bibfnamefont {I.}~\bibnamefont
  {Knez}}, \bibinfo {author} {\bibfnamefont {R.-R.}\ \bibnamefont {Du}}, \ and\
  \bibinfo {author} {\bibfnamefont {G.}~\bibnamefont {Sullivan}},\ }\href
  {\doibase 10.1103/PhysRevLett.107.136603} {\bibfield  {journal} {\bibinfo
  {journal} {Physical Review Letters}\ }\textbf {\bibinfo {volume} {107}},\
  \bibinfo {pages} {136603} (\bibinfo {year} {2011})}\BibitemShut {NoStop}%
\bibitem [{\citenamefont {Suzuki}\ \emph {et~al.}(2013)\citenamefont {Suzuki},
  \citenamefont {Harada}, \citenamefont {Onomitsu},\ and\ \citenamefont
  {Muraki}}]{suzuki_edge_2013}%
  \BibitemOpen
  \bibfield  {author} {\bibinfo {author} {\bibfnamefont {K.}~\bibnamefont
  {Suzuki}}, \bibinfo {author} {\bibfnamefont {Y.}~\bibnamefont {Harada}},
  \bibinfo {author} {\bibfnamefont {K.}~\bibnamefont {Onomitsu}}, \ and\
  \bibinfo {author} {\bibfnamefont {K.}~\bibnamefont {Muraki}},\ }\href
  {\doibase 10.1103/PhysRevB.87.235311} {\bibfield  {journal} {\bibinfo
  {journal} {Physical Review B}\ }\textbf {\bibinfo {volume} {87}},\ \bibinfo
  {pages} {235311} (\bibinfo {year} {2013})}\BibitemShut {NoStop}%
\bibitem [{\citenamefont {Knez}\ \emph {et~al.}(2014)\citenamefont {Knez},
  \citenamefont {Rettner}, \citenamefont {Yang}, \citenamefont {Parkin},
  \citenamefont {Du}, \citenamefont {Du},\ and\ \citenamefont
  {Sullivan}}]{knez_observation_2014}%
  \BibitemOpen
  \bibfield  {author} {\bibinfo {author} {\bibfnamefont {I.}~\bibnamefont
  {Knez}}, \bibinfo {author} {\bibfnamefont {C.~T.}\ \bibnamefont {Rettner}},
  \bibinfo {author} {\bibfnamefont {S.-H.}\ \bibnamefont {Yang}}, \bibinfo
  {author} {\bibfnamefont {S.~S.}\ \bibnamefont {Parkin}}, \bibinfo {author}
  {\bibfnamefont {L.}~\bibnamefont {Du}}, \bibinfo {author} {\bibfnamefont
  {R.-R.}\ \bibnamefont {Du}}, \ and\ \bibinfo {author} {\bibfnamefont
  {G.}~\bibnamefont {Sullivan}},\ }\href {\doibase
  10.1103/PhysRevLett.112.026602} {\bibfield  {journal} {\bibinfo  {journal}
  {Physical Review Letters}\ }\textbf {\bibinfo {volume} {112}},\ \bibinfo
  {pages} {026602} (\bibinfo {year} {2014})}\BibitemShut {NoStop}%
\bibitem [{\citenamefont {Mueller}\ \emph {et~al.}(2015)\citenamefont
  {Mueller}, \citenamefont {Pal}, \citenamefont {Karalic}, \citenamefont
  {Tschirky}, \citenamefont {Charpentier}, \citenamefont {Wegscheider},
  \citenamefont {Ensslin},\ and\ \citenamefont {Ihn}}]{mueller_nonlocal_2015}%
  \BibitemOpen
  \bibfield  {author} {\bibinfo {author} {\bibfnamefont {S.}~\bibnamefont
  {Mueller}}, \bibinfo {author} {\bibfnamefont {A.~N.}\ \bibnamefont {Pal}},
  \bibinfo {author} {\bibfnamefont {M.}~\bibnamefont {Karalic}}, \bibinfo
  {author} {\bibfnamefont {T.}~\bibnamefont {Tschirky}}, \bibinfo {author}
  {\bibfnamefont {C.}~\bibnamefont {Charpentier}}, \bibinfo {author}
  {\bibfnamefont {W.}~\bibnamefont {Wegscheider}}, \bibinfo {author}
  {\bibfnamefont {K.}~\bibnamefont {Ensslin}}, \ and\ \bibinfo {author}
  {\bibfnamefont {T.}~\bibnamefont {Ihn}},\ }\href {\doibase
  10.1103/PhysRevB.92.081303} {\bibfield  {journal} {\bibinfo  {journal}
  {Physical Review B}\ }\textbf {\bibinfo {volume} {92}},\ \bibinfo {pages}
  {081303} (\bibinfo {year} {2015})}\BibitemShut {NoStop}%
\bibitem [{\citenamefont {Du}\ \emph {et~al.}(2015)\citenamefont {Du},
  \citenamefont {Knez}, \citenamefont {Sullivan},\ and\ \citenamefont
  {Du}}]{du_robust_2015}%
  \BibitemOpen
  \bibfield  {author} {\bibinfo {author} {\bibfnamefont {L.}~\bibnamefont
  {Du}}, \bibinfo {author} {\bibfnamefont {I.}~\bibnamefont {Knez}}, \bibinfo
  {author} {\bibfnamefont {G.}~\bibnamefont {Sullivan}}, \ and\ \bibinfo
  {author} {\bibfnamefont {R.-R.}\ \bibnamefont {Du}},\ }\href {\doibase
  10.1103/PhysRevLett.114.096802} {\bibfield  {journal} {\bibinfo  {journal}
  {Physical Review Letters}\ }\textbf {\bibinfo {volume} {114}},\ \bibinfo
  {pages} {096802} (\bibinfo {year} {2015})}\BibitemShut {NoStop}%
\bibitem [{\citenamefont {Beukman}\ \emph {et~al.}(2017)\citenamefont
  {Beukman}, \citenamefont {de~Vries}, \citenamefont {van Veen}, \citenamefont
  {Skolasinski}, \citenamefont {Wimmer}, \citenamefont {Qu}, \citenamefont
  {de~Vries}, \citenamefont {Nguyen}, \citenamefont {Yi}, \citenamefont
  {Kiselev}, \citenamefont {Sokolich}, \citenamefont {Manfra}, \citenamefont
  {Nichele}, \citenamefont {Marcus},\ and\ \citenamefont
  {Kouwenhoven}}]{beukman_spin-orbit_2017}%
  \BibitemOpen
  \bibfield  {author} {\bibinfo {author} {\bibfnamefont {A.~J.~A.}\
  \bibnamefont {Beukman}}, \bibinfo {author} {\bibfnamefont {F.~K.}\
  \bibnamefont {de~Vries}}, \bibinfo {author} {\bibfnamefont {J.}~\bibnamefont
  {van Veen}}, \bibinfo {author} {\bibfnamefont {R.}~\bibnamefont
  {Skolasinski}}, \bibinfo {author} {\bibfnamefont {M.}~\bibnamefont {Wimmer}},
  \bibinfo {author} {\bibfnamefont {F.}~\bibnamefont {Qu}}, \bibinfo {author}
  {\bibfnamefont {D.~T.}\ \bibnamefont {de~Vries}}, \bibinfo {author}
  {\bibfnamefont {B.-M.}\ \bibnamefont {Nguyen}}, \bibinfo {author}
  {\bibfnamefont {W.}~\bibnamefont {Yi}}, \bibinfo {author} {\bibfnamefont
  {A.~A.}\ \bibnamefont {Kiselev}}, \bibinfo {author} {\bibfnamefont
  {M.}~\bibnamefont {Sokolich}}, \bibinfo {author} {\bibfnamefont {M.~J.}\
  \bibnamefont {Manfra}}, \bibinfo {author} {\bibfnamefont {F.}~\bibnamefont
  {Nichele}}, \bibinfo {author} {\bibfnamefont {C.~M.}\ \bibnamefont {Marcus}},
  \ and\ \bibinfo {author} {\bibfnamefont {L.~P.}\ \bibnamefont
  {Kouwenhoven}},\ }\href {\doibase 10.1103/PhysRevB.96.241401} {\bibfield
  {journal} {\bibinfo  {journal} {Physical Review B}\ }\textbf {\bibinfo
  {volume} {96}},\ \bibinfo {pages} {241401} (\bibinfo {year}
  {2017})}\BibitemShut {NoStop}%
\bibitem [{\citenamefont {Nichele}\ \emph {et~al.}(2017)\citenamefont
  {Nichele}, \citenamefont {Kjaergaard}, \citenamefont {Suominen},
  \citenamefont {Skolasinski}, \citenamefont {Wimmer}, \citenamefont {Nguyen},
  \citenamefont {Kiselev}, \citenamefont {Yi}, \citenamefont {Sokolich},
  \citenamefont {Manfra}, \citenamefont {Qu}, \citenamefont {Beukman},
  \citenamefont {Kouwenhoven},\ and\ \citenamefont
  {Marcus}}]{nichele_giant_2017}%
  \BibitemOpen
  \bibfield  {author} {\bibinfo {author} {\bibfnamefont {F.}~\bibnamefont
  {Nichele}}, \bibinfo {author} {\bibfnamefont {M.}~\bibnamefont {Kjaergaard}},
  \bibinfo {author} {\bibfnamefont {H.~J.}\ \bibnamefont {Suominen}}, \bibinfo
  {author} {\bibfnamefont {R.}~\bibnamefont {Skolasinski}}, \bibinfo {author}
  {\bibfnamefont {M.}~\bibnamefont {Wimmer}}, \bibinfo {author} {\bibfnamefont
  {B.-M.}\ \bibnamefont {Nguyen}}, \bibinfo {author} {\bibfnamefont {A.~A.}\
  \bibnamefont {Kiselev}}, \bibinfo {author} {\bibfnamefont {W.}~\bibnamefont
  {Yi}}, \bibinfo {author} {\bibfnamefont {M.}~\bibnamefont {Sokolich}},
  \bibinfo {author} {\bibfnamefont {M.~J.}\ \bibnamefont {Manfra}}, \bibinfo
  {author} {\bibfnamefont {F.}~\bibnamefont {Qu}}, \bibinfo {author}
  {\bibfnamefont {A.~J.}\ \bibnamefont {Beukman}}, \bibinfo {author}
  {\bibfnamefont {L.~P.}\ \bibnamefont {Kouwenhoven}}, \ and\ \bibinfo {author}
  {\bibfnamefont {C.~M.}\ \bibnamefont {Marcus}},\ }\href {\doibase
  10.1103/PhysRevLett.118.016801} {\bibfield  {journal} {\bibinfo  {journal}
  {Physical Review Letters}\ }\textbf {\bibinfo {volume} {118}},\ \bibinfo
  {pages} {016801} (\bibinfo {year} {2017})}\BibitemShut {NoStop}%
\bibitem [{\citenamefont {Karalic}\ \emph {et~al.}(2017)\citenamefont
  {Karalic}, \citenamefont {Mittag}, \citenamefont {Tschirky}, \citenamefont
  {Wegscheider}, \citenamefont {Ensslin},\ and\ \citenamefont
  {Ihn}}]{karalic_lateral_2017}%
  \BibitemOpen
  \bibfield  {author} {\bibinfo {author} {\bibfnamefont {M.}~\bibnamefont
  {Karalic}}, \bibinfo {author} {\bibfnamefont {C.}~\bibnamefont {Mittag}},
  \bibinfo {author} {\bibfnamefont {T.}~\bibnamefont {Tschirky}}, \bibinfo
  {author} {\bibfnamefont {W.}~\bibnamefont {Wegscheider}}, \bibinfo {author}
  {\bibfnamefont {K.}~\bibnamefont {Ensslin}}, \ and\ \bibinfo {author}
  {\bibfnamefont {T.}~\bibnamefont {Ihn}},\ }\href {\doibase
  10.1103/PhysRevLett.118.206801} {\bibfield  {journal} {\bibinfo  {journal}
  {Physical Review Letters}\ }\textbf {\bibinfo {volume} {118}},\ \bibinfo
  {pages} {206801} (\bibinfo {year} {2017})}\BibitemShut {NoStop}%
\bibitem [{\citenamefont {Zakharova}\ \emph {et~al.}(2002)\citenamefont
  {Zakharova}, \citenamefont {Yen},\ and\ \citenamefont
  {Chao}}]{zakharova_strain-induced_2002}%
  \BibitemOpen
  \bibfield  {author} {\bibinfo {author} {\bibfnamefont {A.}~\bibnamefont
  {Zakharova}}, \bibinfo {author} {\bibfnamefont {S.~T.}\ \bibnamefont {Yen}},
  \ and\ \bibinfo {author} {\bibfnamefont {K.~A.}\ \bibnamefont {Chao}},\
  }\href {\doibase 10.1103/PhysRevB.66.085312} {\bibfield  {journal} {\bibinfo
  {journal} {Physical Review B}\ }\textbf {\bibinfo {volume} {66}},\ \bibinfo
  {pages} {085312} (\bibinfo {year} {2002})}\BibitemShut {NoStop}%
\bibitem [{\citenamefont {Li}\ \emph {et~al.}(2009)\citenamefont {Li},
  \citenamefont {Yang},\ and\ \citenamefont {Chang}}]{li_spin_2009}%
  \BibitemOpen
  \bibfield  {author} {\bibinfo {author} {\bibfnamefont {J.}~\bibnamefont
  {Li}}, \bibinfo {author} {\bibfnamefont {W.}~\bibnamefont {Yang}}, \ and\
  \bibinfo {author} {\bibfnamefont {K.}~\bibnamefont {Chang}},\ }\href
  {\doibase 10.1103/PhysRevB.80.035303} {\bibfield  {journal} {\bibinfo
  {journal} {Physical Review B}\ }\textbf {\bibinfo {volume} {80}},\ \bibinfo
  {pages} {035303} (\bibinfo {year} {2009})}\BibitemShut {NoStop}%
\bibitem [{\citenamefont {Hu}\ \emph {et~al.}(2016)\citenamefont {Hu},
  \citenamefont {Liu}, \citenamefont {Xu}, \citenamefont {Zhang},\ and\
  \citenamefont {Zhou}}]{Hu_electric_2016}%
  \BibitemOpen
  \bibfield  {author} {\bibinfo {author} {\bibfnamefont {L.-H.}\ \bibnamefont
  {Hu}}, \bibinfo {author} {\bibfnamefont {C.-X.}\ \bibnamefont {Liu}},
  \bibinfo {author} {\bibfnamefont {D.-H.}\ \bibnamefont {Xu}}, \bibinfo
  {author} {\bibfnamefont {F.-C.}\ \bibnamefont {Zhang}}, \ and\ \bibinfo
  {author} {\bibfnamefont {Y.}~\bibnamefont {Zhou}},\ }\href {\doibase
  10.1103/PhysRevB.94.045317} {\bibfield  {journal} {\bibinfo  {journal}
  {Physical Review B}\ }\textbf {\bibinfo {volume} {94}},\ \bibinfo {pages}
  {045317} (\bibinfo {year} {2016})}\BibitemShut {NoStop}%
\bibitem [{\citenamefont {Movva}\ \emph {et~al.}(2017)\citenamefont {Movva},
  \citenamefont {Fallahazad}, \citenamefont {Kim}, \citenamefont {Larentis},
  \citenamefont {Taniguchi}, \citenamefont {Watanabe}, \citenamefont
  {Banerjee},\ and\ \citenamefont {Tutuc}}]{movva_density-dependent_2017}%
  \BibitemOpen
  \bibfield  {author} {\bibinfo {author} {\bibfnamefont {H.~C.}\ \bibnamefont
  {Movva}}, \bibinfo {author} {\bibfnamefont {B.}~\bibnamefont {Fallahazad}},
  \bibinfo {author} {\bibfnamefont {K.}~\bibnamefont {Kim}}, \bibinfo {author}
  {\bibfnamefont {S.}~\bibnamefont {Larentis}}, \bibinfo {author}
  {\bibfnamefont {T.}~\bibnamefont {Taniguchi}}, \bibinfo {author}
  {\bibfnamefont {K.}~\bibnamefont {Watanabe}}, \bibinfo {author}
  {\bibfnamefont {S.~K.}\ \bibnamefont {Banerjee}}, \ and\ \bibinfo {author}
  {\bibfnamefont {E.}~\bibnamefont {Tutuc}},\ }\href {\doibase
  10.1103/PhysRevLett.118.247701} {\bibfield  {journal} {\bibinfo  {journal}
  {Physical Review Letters}\ }\textbf {\bibinfo {volume} {118}},\ \bibinfo
  {pages} {247701} (\bibinfo {year} {2017})}\BibitemShut {NoStop}%
\bibitem [{\citenamefont {Larentis}\ \emph {et~al.}(2018)\citenamefont
  {Larentis}, \citenamefont {Movva}, \citenamefont {Fallahazad}, \citenamefont
  {Kim}, \citenamefont {Behroozi}, \citenamefont {Taniguchi}, \citenamefont
  {Watanabe}, \citenamefont {Banerjee},\ and\ \citenamefont
  {Tutuc}}]{larentis_large_2018}%
  \BibitemOpen
  \bibfield  {author} {\bibinfo {author} {\bibfnamefont {S.}~\bibnamefont
  {Larentis}}, \bibinfo {author} {\bibfnamefont {H.~C.~P.}\ \bibnamefont
  {Movva}}, \bibinfo {author} {\bibfnamefont {B.}~\bibnamefont {Fallahazad}},
  \bibinfo {author} {\bibfnamefont {K.}~\bibnamefont {Kim}}, \bibinfo {author}
  {\bibfnamefont {A.}~\bibnamefont {Behroozi}}, \bibinfo {author}
  {\bibfnamefont {T.}~\bibnamefont {Taniguchi}}, \bibinfo {author}
  {\bibfnamefont {K.}~\bibnamefont {Watanabe}}, \bibinfo {author}
  {\bibfnamefont {S.~K.}\ \bibnamefont {Banerjee}}, \ and\ \bibinfo {author}
  {\bibfnamefont {E.}~\bibnamefont {Tutuc}},\ }\href {\doibase
  10.1103/PhysRevB.97.201407} {\bibfield  {journal} {\bibinfo  {journal}
  {Physical Review B}\ }\textbf {\bibinfo {volume} {97}},\ \bibinfo {pages}
  {201407} (\bibinfo {year} {2018})}\BibitemShut {NoStop}%
\bibitem [{\citenamefont {Pisoni}\ \emph {et~al.}(2018)\citenamefont {Pisoni},
  \citenamefont {Kormányos}, \citenamefont {Brooks}, \citenamefont {Lei},
  \citenamefont {Back}, \citenamefont {Eich}, \citenamefont {Overweg},
  \citenamefont {Lee}, \citenamefont {Rickhaus}, \citenamefont {Watanabe},
  \citenamefont {Taniguchi}, \citenamefont {Imamoglu}, \citenamefont {Burkard},
  \citenamefont {Ihn},\ and\ \citenamefont
  {Ensslin}}]{pisoni_interactions_2018}%
  \BibitemOpen
  \bibfield  {author} {\bibinfo {author} {\bibfnamefont {R.}~\bibnamefont
  {Pisoni}}, \bibinfo {author} {\bibfnamefont {A.}~\bibnamefont {Kormányos}},
  \bibinfo {author} {\bibfnamefont {M.}~\bibnamefont {Brooks}}, \bibinfo
  {author} {\bibfnamefont {Z.}~\bibnamefont {Lei}}, \bibinfo {author}
  {\bibfnamefont {P.}~\bibnamefont {Back}}, \bibinfo {author} {\bibfnamefont
  {M.}~\bibnamefont {Eich}}, \bibinfo {author} {\bibfnamefont {H.}~\bibnamefont
  {Overweg}}, \bibinfo {author} {\bibfnamefont {Y.}~\bibnamefont {Lee}},
  \bibinfo {author} {\bibfnamefont {P.}~\bibnamefont {Rickhaus}}, \bibinfo
  {author} {\bibfnamefont {K.}~\bibnamefont {Watanabe}}, \bibinfo {author}
  {\bibfnamefont {T.}~\bibnamefont {Taniguchi}}, \bibinfo {author}
  {\bibfnamefont {A.}~\bibnamefont {Imamoglu}}, \bibinfo {author}
  {\bibfnamefont {G.}~\bibnamefont {Burkard}}, \bibinfo {author} {\bibfnamefont
  {T.}~\bibnamefont {Ihn}}, \ and\ \bibinfo {author} {\bibfnamefont
  {K.}~\bibnamefont {Ensslin}},\ }\href {http://arxiv.org/abs/1806.06402}
  {\bibfield  {journal} {\bibinfo  {journal} {arXiv:1806.06402 [cond-mat]}\ }
  (\bibinfo {year} {2018})},\ \bibinfo {note} {arXiv: 1806.06402}\BibitemShut
  {NoStop}%
\bibitem [{\citenamefont {Karalic}\ \emph {et~al.}(2016)\citenamefont
  {Karalic}, \citenamefont {Mueller}, \citenamefont {Mittag}, \citenamefont
  {Pakrouski}, \citenamefont {Wu}, \citenamefont {Soluyanov}, \citenamefont
  {Troyer}, \citenamefont {Tschirky}, \citenamefont {Wegscheider},
  \citenamefont {Ensslin},\ and\ \citenamefont
  {Ihn}}]{karalic_experimental_2016}%
  \BibitemOpen
  \bibfield  {author} {\bibinfo {author} {\bibfnamefont {M.}~\bibnamefont
  {Karalic}}, \bibinfo {author} {\bibfnamefont {S.}~\bibnamefont {Mueller}},
  \bibinfo {author} {\bibfnamefont {C.}~\bibnamefont {Mittag}}, \bibinfo
  {author} {\bibfnamefont {K.}~\bibnamefont {Pakrouski}}, \bibinfo {author}
  {\bibfnamefont {Q.}~\bibnamefont {Wu}}, \bibinfo {author} {\bibfnamefont
  {A.~A.}\ \bibnamefont {Soluyanov}}, \bibinfo {author} {\bibfnamefont
  {M.}~\bibnamefont {Troyer}}, \bibinfo {author} {\bibfnamefont
  {T.}~\bibnamefont {Tschirky}}, \bibinfo {author} {\bibfnamefont
  {W.}~\bibnamefont {Wegscheider}}, \bibinfo {author} {\bibfnamefont
  {K.}~\bibnamefont {Ensslin}}, \ and\ \bibinfo {author} {\bibfnamefont
  {T.}~\bibnamefont {Ihn}},\ }\href {\doibase 10.1103/PhysRevB.94.241402}
  {\bibfield  {journal} {\bibinfo  {journal} {Physical Review B}\ }\textbf
  {\bibinfo {volume} {94}},\ \bibinfo {pages} {241402} (\bibinfo {year}
  {2016})}\BibitemShut {NoStop}%
  \bibitem{supp} For more details refer to the Supplemental Material at [URL will be inserted by publisher].  %supp
  \bibitem{note1} For pure InAs, the ratio of Zeeman and cyclotron energies is ${\sim}0.26$ for $g = 15$ and $m^* = 0.036 \times m_0$. Due to the presence of hole-like Landau levels, 
  the spectrum is denser than expected for pure InAs electron systems.  %note1 
\bibitem [{\citenamefont {Novoselov}\ \emph {et~al.}(2005)\citenamefont
  {Novoselov}, \citenamefont {Geim}, \citenamefont {Morozov}, \citenamefont
  {Jiang}, \citenamefont {Katsnelson}, \citenamefont {Grigorieva},
  \citenamefont {Dubonos},\ and\ \citenamefont
  {Firsov}}]{novoselov_two-dimensional_2005}%
  \BibitemOpen
  \bibfield  {author} {\bibinfo {author} {\bibfnamefont {K.~S.}\ \bibnamefont
  {Novoselov}}, \bibinfo {author} {\bibfnamefont {A.~K.}\ \bibnamefont {Geim}},
  \bibinfo {author} {\bibfnamefont {S.~V.}\ \bibnamefont {Morozov}}, \bibinfo
  {author} {\bibfnamefont {D.}~\bibnamefont {Jiang}}, \bibinfo {author}
  {\bibfnamefont {M.~I.}\ \bibnamefont {Katsnelson}}, \bibinfo {author}
  {\bibfnamefont {I.~V.}\ \bibnamefont {Grigorieva}}, \bibinfo {author}
  {\bibfnamefont {S.~V.}\ \bibnamefont {Dubonos}}, \ and\ \bibinfo {author}
  {\bibfnamefont {A.~A.}\ \bibnamefont {Firsov}},\ }\href {\doibase
  10.1038/nature04233} {\bibfield  {journal} {\bibinfo  {journal} {Nature}\
  }\textbf {\bibinfo {volume} {438}},\ \bibinfo {pages} {197} (\bibinfo {year}
  {2005})}\BibitemShut {NoStop}%
\bibitem [{\citenamefont {Taskin}\ and\ \citenamefont
  {Ando}(2011)}]{taskin_berry_2011}%
  \BibitemOpen
  \bibfield  {author} {\bibinfo {author} {\bibfnamefont {A.~A.}\ \bibnamefont
  {Taskin}}\ and\ \bibinfo {author} {\bibfnamefont {Y.}~\bibnamefont {Ando}},\
  }\href {\doibase 10.1103/PhysRevB.84.035301} {\bibfield  {journal} {\bibinfo
  {journal} {Physical Review B}\ }\textbf {\bibinfo {volume} {84}},\ \bibinfo
  {pages} {035301} (\bibinfo {year} {2011})}\BibitemShut {NoStop}%
\bibitem [{\citenamefont {Wright}\ and\ \citenamefont
  {McKenzie}(2013)}]{wright_quantum_2013}%
  \BibitemOpen
  \bibfield  {author} {\bibinfo {author} {\bibfnamefont {A.~R.}\ \bibnamefont
  {Wright}}\ and\ \bibinfo {author} {\bibfnamefont {R.~H.}\ \bibnamefont
  {McKenzie}},\ }\href {\doibase 10.1103/PhysRevB.87.085411} {\bibfield
  {journal} {\bibinfo  {journal} {Physical Review B}\ }\textbf {\bibinfo
  {volume} {87}},\ \bibinfo {pages} {085411} (\bibinfo {year}
  {2013})}\BibitemShut {NoStop}%
\bibitem [{\citenamefont {Kuntsevich}\ \emph {et~al.}(2018)\citenamefont
  {Kuntsevich}, \citenamefont {Shupletsov},\ and\ \citenamefont
  {Minkov}}]{kuntsevich_simple_2018}%
  \BibitemOpen
  \bibfield  {author} {\bibinfo {author} {\bibfnamefont {A.~Y.}\ \bibnamefont
  {Kuntsevich}}, \bibinfo {author} {\bibfnamefont {A.~V.}\ \bibnamefont
  {Shupletsov}}, \ and\ \bibinfo {author} {\bibfnamefont {G.~M.}\ \bibnamefont
  {Minkov}},\ }\href {\doibase 10.1103/PhysRevB.97.195431} {\bibfield
  {journal} {\bibinfo  {journal} {Physical Review B}\ }\textbf {\bibinfo
  {volume} {97}},\ \bibinfo {pages} {195431} (\bibinfo {year}
  {2018})}\BibitemShut {NoStop}%
\bibitem [{\citenamefont {König}\ \emph {et~al.}(2007)\citenamefont {König},
  \citenamefont {Wiedmann}, \citenamefont {Brüne}, \citenamefont {Roth},
  \citenamefont {Buhmann}, \citenamefont {Molenkamp}, \citenamefont {Qi},\ and\
  \citenamefont {Zhang}}]{konig_quantum_2007}%
  \BibitemOpen
  \bibfield  {author} {\bibinfo {author} {\bibfnamefont {M.}~\bibnamefont
  {K\"onig}}, \bibinfo {author} {\bibfnamefont {S.}~\bibnamefont {Wiedmann}},
  \bibinfo {author} {\bibfnamefont {C.}~\bibnamefont {Br\"une}}, \bibinfo
  {author} {\bibfnamefont {A.}~\bibnamefont {Roth}}, \bibinfo {author}
  {\bibfnamefont {H.}~\bibnamefont {Buhmann}}, \bibinfo {author} {\bibfnamefont
  {L.~W.}\ \bibnamefont {Molenkamp}}, \bibinfo {author} {\bibfnamefont {X.-L.}\
  \bibnamefont {Qi}}, \ and\ \bibinfo {author} {\bibfnamefont {S.-C.}\
  \bibnamefont {Zhang}},\ }\href {\doibase 10.1126/science.1148047} {\bibfield
  {journal} {\bibinfo  {journal} {Science}\ }\textbf {\bibinfo {volume}
  {318}},\ \bibinfo {pages} {766} (\bibinfo {year} {2007})}\BibitemShut
  {NoStop}%
\bibitem [{\citenamefont {Qian}\ \emph {et~al.}(2014)\citenamefont {Qian},
  \citenamefont {Liu}, \citenamefont {Fu},\ and\ \citenamefont
  {Li}}]{qian_quantum_2014}%
  \BibitemOpen
  \bibfield  {author} {\bibinfo {author} {\bibfnamefont {X.}~\bibnamefont
  {Qian}}, \bibinfo {author} {\bibfnamefont {J.}~\bibnamefont {Liu}}, \bibinfo
  {author} {\bibfnamefont {L.}~\bibnamefont {Fu}}, \ and\ \bibinfo {author}
  {\bibfnamefont {J.}~\bibnamefont {Li}},\ }\href {\doibase
  10.1126/science.1256815} {\bibfield  {journal} {\bibinfo  {journal}
  {Science}\ }\textbf {\bibinfo {volume} {346}},\ \bibinfo {pages} {1344}
  (\bibinfo {year} {2014})}\BibitemShut {NoStop}%
\bibitem [{\citenamefont {Wu}\ \emph {et~al.}(2018)\citenamefont {Wu},
  \citenamefont {Fatemi}, \citenamefont {Gibson}, \citenamefont {Watanabe},
  \citenamefont {Taniguchi}, \citenamefont {Cava},\ and\ \citenamefont
  {Jarillo-Herrero}}]{wu_observation_2018}%
  \BibitemOpen
  \bibfield  {author} {\bibinfo {author} {\bibfnamefont {S.}~\bibnamefont
  {Wu}}, \bibinfo {author} {\bibfnamefont {V.}~\bibnamefont {Fatemi}}, \bibinfo
  {author} {\bibfnamefont {Q.~D.}\ \bibnamefont {Gibson}}, \bibinfo {author}
  {\bibfnamefont {K.}~\bibnamefont {Watanabe}}, \bibinfo {author}
  {\bibfnamefont {T.}~\bibnamefont {Taniguchi}}, \bibinfo {author}
  {\bibfnamefont {R.~J.}\ \bibnamefont {Cava}}, \ and\ \bibinfo {author}
  {\bibfnamefont {P.}~\bibnamefont {Jarillo-Herrero}},\ }\href {\doibase
  10.1126/science.aan6003} {\bibfield  {journal} {\bibinfo  {journal}
  {Science}\ }\textbf {\bibinfo {volume} {359}},\ \bibinfo {pages} {76}
  (\bibinfo {year} {2018})}\BibitemShut {NoStop}%
\end{thebibliography}
